\documentclass[twocolumn,amsmath,amssymb]{revtex4}
\usepackage{graphicx}
\begin{document}

\title{Guided-wave Brillouin scattering in air}

\author{William H. Renninger$^{*}$, Ryan O. Behunin, and Peter T. Rakich}
\affiliation{Department of Applied Physics, Yale University, New Haven, Connecticut 06520\\
$^*$Corresponding author: william.renninger@yale.edu}

\begin{abstract}
Here we identify a new form of optomechanical coupling in gas-filled hollow-core fibers. Stimulated forward Brillouin scattering is observed in air in the core of a photonic bandgap fiber.  A single resonance is observed at 35 MHz, which corresponds to the first excited axial-radial acoustic mode in the air-filled core.  The linewidth and coupling strengths are determined by the acoustic loss and electrostrictive coupling in air, respectively.  A simple analytical model, refined by numerical simulations, is developed that accurately predicts the Brillouin coupling strength and frequency from the gas and fiber parameters. Since this form of Brillouin coupling depends strongly on both the acoustic and dispersive optical properties of the gas within the fiber, this new type of optomechanical interaction is highly tailorable.   These results allow for forward Brillouin spectroscopy in dilute gases, could be useful for sensing and will present a power and noise limitation for certain applications.
\end{abstract}

\maketitle

\begin{figure}[htb]
\centerline{
\includegraphics[width=8.0cm]{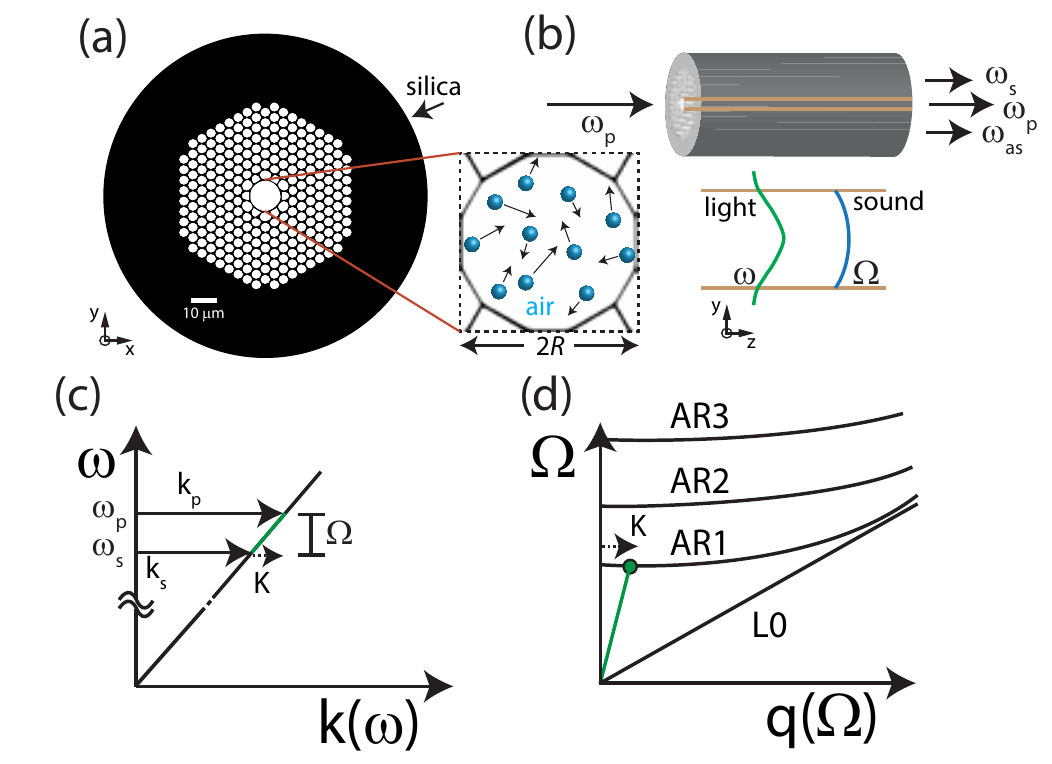}}
\caption{Hollow-core fiber (a) cross section, (b) Brillouin frequencies of interest in the HC-PBF, indicated as inputs and outputs, (c) optical dispersion curve, and (d) the acoustic mode dispersion which satisfies phase matching.  The inset represents the vibrating air in the core of the fiber.}
\label{introfig}
\end{figure}

Hollow-core photonic bandgap fibers (HC-PBF) are unique for their ability to guide light in air through Bragg reflection from a periodic silica matrix that forms the waveguide cladding \cite{Hess2000,Cregan1999,Russell2003} (Fig. \ref{introfig}(a)). In comparison to conventional silica (step-index) fibers, Bragg guidance in HC-PBF drastically reduces the nonlinear interactions with silica and increases the power handling to permit new forms of high power laser delivery \cite{Ouzounov2003}, pulse compression \cite{nisoli1996}, and light sources \cite{Couny2007}.  Conversely, the introduction of atomic vapors within these hollow-core fibers produces sustained photon-atom interactions over unprecedented length scales, which enables new light sources for both classical \cite{Russell2014} and quantum \cite{Zhong2007,Ghosh2006,Light2007,Vogl2014,Sprague2014} applications. While electronic nonlinearities are widely studied in such fibers \cite{Hess2000,Cregan1999,Russell2003}, comparatively little is known of their acousto-optic (or optomechanical) interactions, and the dynamics, and noise that they can produce.  Only recently has coupling between MHz elastic waves within the silica matrix been quantified \cite{Renninger2016}, and identified as a source of noise in quantum optics \cite{Zhong2015,Renninger2016}.

In this paper, we show that photon-phonon coupling mediated by air (gasses) within the hollow core of the fiber constitutes a much larger and perhaps more tailorable form of optomechanical coupling.  Through a combination of theory and experiment, we show that the hollow core of the fiber acts as a conduit for guided acoustic waves in air.  Optical waves that are guided in this same region produce strong photo-acoustic coupling to these sound waves, yielding appreciable forward-Brillouin coupling at MHz frequencies.  Using precision spectroscopy methods, we identify a single air-mediated Brillouin resonance at 35 MHz with 20 times stronger photo-acoustic coupling than is produced by elastic waves in the silica cladding alone. We show that the strength, frequency, and character of this Brillouin resonance is explained by the properties of the gas filling the core and the dimension of the hollow-core fiber.  Since this form of Brillouin coupling depends strongly on both the acoustic and dispersive optical properties of the gas within the fiber, this new type of optomechanical interaction is highly tailorable. More generally, gas optomechanics may lead to new forms of spectroscopy, will be valuable for sensing, and will present a power and noise limitation for certain applications.  In what follows, we identify the optomechanical properties of air in HC-PBFs.

Through stimulated Brillouin scattering processes, interfering optical waves produce time-varying optical forces that drive the excitation of acoustic waves. These elastic waves scatter light to new frequencies through dynamical changes in the dielectric response. Depending on the type of deformation and the medium response, light can be scattered to optical modes with similar or dissimilar polarization states. These distinct scattering processes are termed intra-polarization and inter-polarization scattering respectively.  It is important to note that while elastic solids can mediate both intra- and inter-polarization scattering (e.g. see Ref. \cite{Shelby1985a}), inter-polarization scattering is generally forbidden in gasses.  Both scattering processes occur in solids because elastic deformation of a solid produces optical birefringence (through the photoelastic effect) that readily changes the polarization of light. By contrast, pressure waves do not produce birefringence in typical gases (e.g. nonchiral), meaning that inter-polarization scattering is forbidden.

In what follows, we consider forward stimulated Brillouin scattering (SBS) processes \cite{Shelby1985a,Kang2009,Dainese2006,Beugnot2007,Kang2008a,Wang2011,Rakich2012,shin2013,Renninger2016}, not to be confused with more widely studied backwards stimulated Brillouin scattering processes \cite{boyd}.  Through forward-SBS, co-propagating pump and Stokes waves of frequencies $\omega_p$ and $\omega_s$ are coupled through parametrically generated phonons of frequency $\Omega=\omega_p-\omega_s$, as indicated by Fig. \ref{introfig}(b).  Coupling is mediated by guided acoustic phonons that satisfy the phase matching condition  $K(\Omega)=k(\omega_p)-k(\omega_s)$, where $K(\Omega)$ and $k(\omega)$ are the acoustic and optical dispersion relations of the type sketched in Fig. \ref{introfig}(c-d). An analogous set of conditions apply for the anti-Stokes process.

In contrast to backward-SBS processes, the phase matching condition for forward-SBS is not satisfied by elastic waves in bulk media, and typically requires wave-guidance for both light and sound \cite{Shelby1985a,Kang2009,Dainese2006,Beugnot2007,Kang2008a,Wang2011,Rakich2012,shin2013,Renninger2016}. Due to the small wave-vector mismatch between the pump and Stokes fields, only guided-acoustic phonons with nonzero frequencies for $K = 0$ can mediate forward Brillouin coupling. Within the HC-PBF of Fig. \ref{introfig}, this phase matching condition is only satisfied by guided acoustic waves that exhibit a frequency cutoff, labeled with mode index AR1-ARN in Fig. \ref{introfig}(d).

\begin{figure}[htb]
\centerline{
\includegraphics[width=8.0cm]{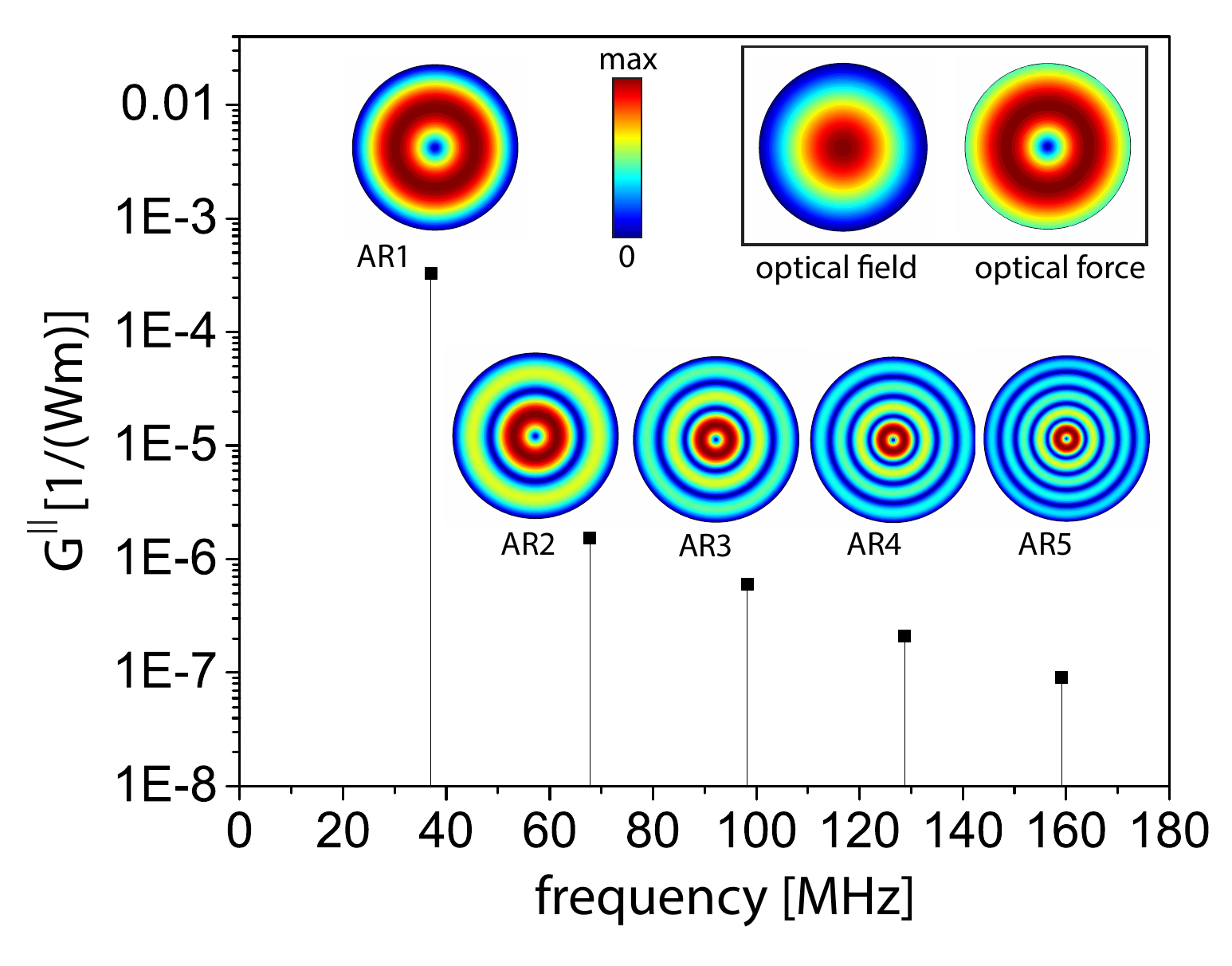}}
\caption{Calculated forward-SBS gain with the magnitude of the acoustic displacement field plotted by the respective gain peak. Inset: Gaussian electric field field and resultant electrostrictive force distribution.  Calculations assume an inner diameter of $5.65$ $\mu$m, $Q=8.75$, $\lambda=1550$ nm, $n=1.0003$ for air, the air density is $1.2754$ kg/m$^3$, and the pressure-wave speed is 343 m/s.}
\label{fig4theory}
\end{figure}

We will analyze this problem using both (1) an approximate analytical model and (2) a full vectorial numerical model of Brillouin coupling.  We begin with an analytical model. The Brillouin coupling strength ($G$) is determined by the overlap of the optical and acoustic fields, as well  as by electrostrictive coupling parameters. By treating the inner glass boundary of the fiber as a hard reflecting boundary for sound, one finds a set of guided acoustic mode solutions (for further details see Supplement 1).  Using the formalism described in Refs. \cite{qiu2013stimulated,RakichOE2,Renninger2016}, one can show that the Brillouin gain coefficient associated with intra- and inter- polarization scattering, denoted as $G^{\parallel}$ and $G^{\perp}$ respectively, stems from the overlap between the dipole (or electrostrictive) forces that act to compress the gas, and the acoustic modes within the hollow core.

Using an approximation for the acoustic profiles in terms of Bessel functions and a Gaussian approximation for the optical profile, one finds
\begin{equation}
G^{\parallel}_i=\frac{\pi Q \xi_i n_g^2(n^2-1)^2}{\lambda X_{1i}^2 v^2 n^4 c \rho R^2}\quad \text{and} \quad G^{\perp}_i=0.
\label{ideagasG}
\end{equation}
Here $n$ is the refractive index, $n_g$ is the group index,  $\rho$ is the mass density, $v$ is the sound velocity, $R$ is the core radius, $c$ is the speed of light, $Q$ is the acoustic quality factor, $X_{1i}$ is the $i$th zero of the first Bessel function of the first kind ($J_1$), and $\xi_i$ is a nondimensional coupling constant with the first five values (for AR1-AR5, e.g.) given by $\xi=\{1.085,.017,.014,.00832,.00548...\}$.  The Bessel zeros arise from the boundary condition which constrains the radial displacement to zero at the radius of the hollow core.  This same condition determines the acoustic resonance frequencies as $\Omega_i=X_{1i}v/R$; further details can be found in Supplement 1. Note that while the excited axial-radial modes AR1-ARN satisfy the phase-matching conditions for forward Brillouin scattering, the fundamental longitudinal mode ($L0$) cannot (Fig. \ref{introfig}).

The gain in air for the first five resonances is plotted in Fig. \ref{fig4theory} along with the optical field and electrostrictive force distributions. The overlap integral of the first mode dominates over the higher order modes because this mode profile is well-matched to the optical forcing profile. The higher-order modes have alternating directions of displacement, which when forced optically in only one direction, results in a reduced coupling strength. Consequently, the gain for the first mode (AR1) is more than 200 times greater than it is for the next higher-order mode.  Therefore, the simple analytical model predicts a single high-gain resonance at the same polarization as the driving beams (intra-polarization) at a frequency of $\sim35$ MHz, with a coupling strength of $G^{AR1}\sim10^{-3}$ W$^{-1}$m$^{-1}$ and a linewidth consistent with the acoustic loss of air at this frequency (e.g. $\sim3$ MHz \cite{Bond1992}).

\begin{figure}[htb]
\centerline{
\includegraphics[width=8.0cm]{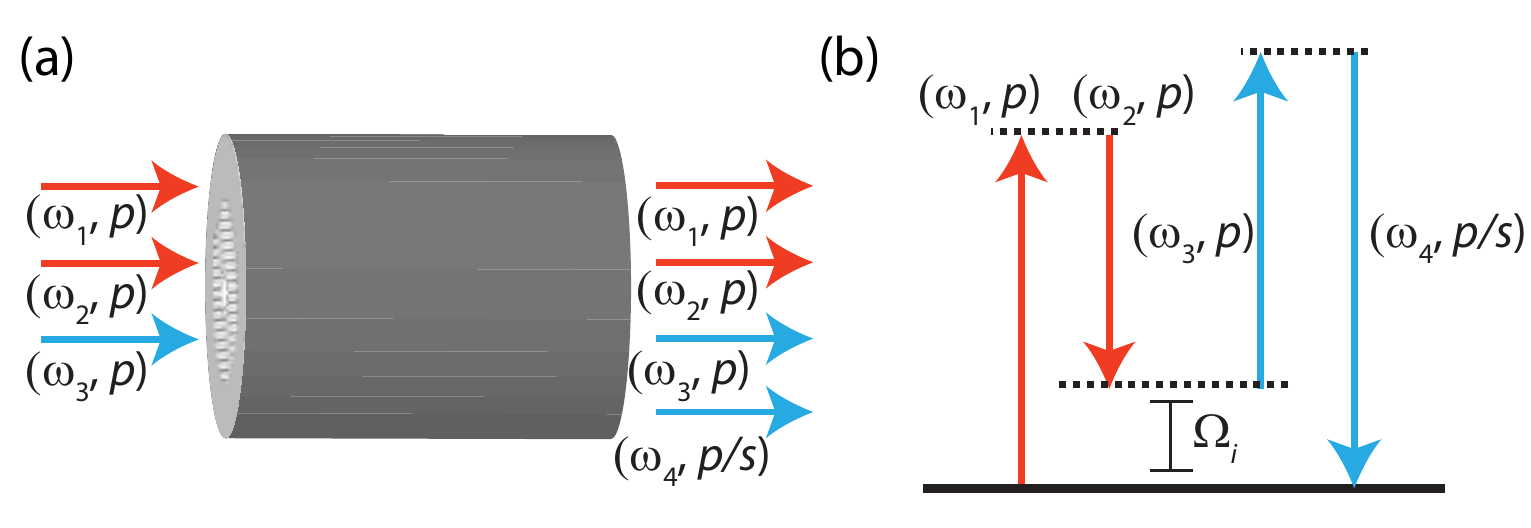}}
\caption{(a) Hollow-core fiber with input and output fields indicated and (b) energy-level diagram for the phonon mediated four-wave mixing process described in this work (both inter or intra- polarized versions).  Each frequency is indicated with a number and a polarization in parenthesis. Note that the energy-level diagram represents one specific anti-Stokes interaction; the corresponding Stokes process occurs simultaneously.}
\label{PHFWMfig}
\end{figure}

We use the two color pump-probe technique of Refs. \cite{Tang1,Fellegara1997,shin2013,Renninger2016} to determine the strength and character of the Brillouin interaction.   The two pump-fields (red), $\omega_1$ and $\omega_2$, are synthesized from a monochromatic laser using an intensity modulator (Fig. \ref{PHFWMfig}).  The modulation frequency is swept through the Brillouin-active resonances producing a resonant excitation of phonons in the HC-PBF. A continuous-wave probe beam (blue), $\omega_3$, is simultaneously injected into the fiber, to permit detection of the excited phonons.  Through this process, a new frequency ($\omega_4$) is detected at the output of the fiber.  By polarization resolving the four frequencies we can separately measure $G^{\parallel}$ and $G^{\perp}$ as a function of acoustic frequency.  In other words, we can quantify inter- and intra-polarization coupling strengths and compare with theory.  In addition, a frequency independent background due to Kerr four-wave mixing is expected.  For further details, see Supplement 1.

\begin{figure}[htb]
\centerline{
\includegraphics[width=8.0cm]{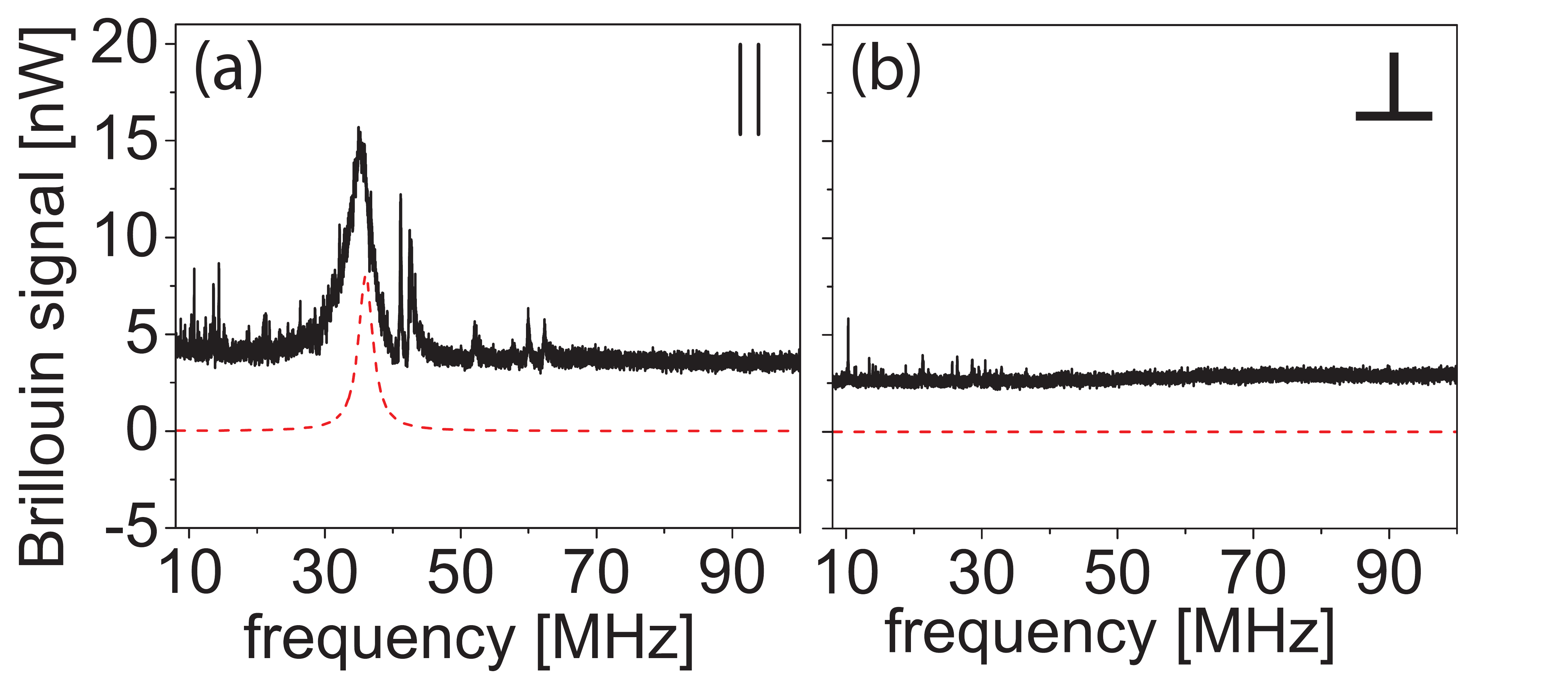}}
\caption{Stimulated Brillouin gain measurement for (a) intra-polarization and (b) inter-polarization scattering (solid black line) along with the theoretical prediction for air coupling, neglecting Kerr nonlinearity and the narrow resonances from the elastic modes in the silica (dashed red line).  The acoustic signature is absent from the inter-polarization measurement as expected for acousto-optic coupling with gases.}
\label{fig2dat}
\end{figure}

Detection of the co-polarized scattered light field versus drive frequency reveals a single 4-MHz linewidth Brillouin-active mode at 35 MHz, as seen in Fig \ref{fig2dat}(a). This Brillouin signature is absent from the orthogonal polarization (Fig. \ref{fig2dat}(b)).  Therefore, the intra-polarization scattering from this resonance is consistent with Brillouin coupling to air.  Multiple narrower peaks are also observed at various frequencies in both polarizations which are consistent with Brillouin scattering from the elastic modes supported by the fiber's silica micro-structure, as described in  Ref. \cite{Renninger2016}. All of these Brillouin signatures sit atop a nonzero background that results from Kerr nonlinearities.

From these spectra, the nonlinear Brillouin gain, $G^{AR1}$, can be determined using the known optical powers, fiber length, and coupling factor (for further details see Supplement 1).  The peak power of the 35-MHz peak is given by 15 nW.  The gain after subtracting the power contribution from Kerr four-wave mixing is then given by $G^{AR1}=9\times10^{-4}$ W$^{-1}$m$^{-1}$.  Therefore, the polarization, frequency, gain and linewidth are all consistent with guided Brillouin scattering in air.  Quantitatively, the forward-FBS gain given by experiment is three times larger than the prediction from the simple theoretical model presented above, which neglects the experimental complexity of the air and silica cladding.

\begin{figure}[htb]
\centerline{
\includegraphics[width=8.0cm]{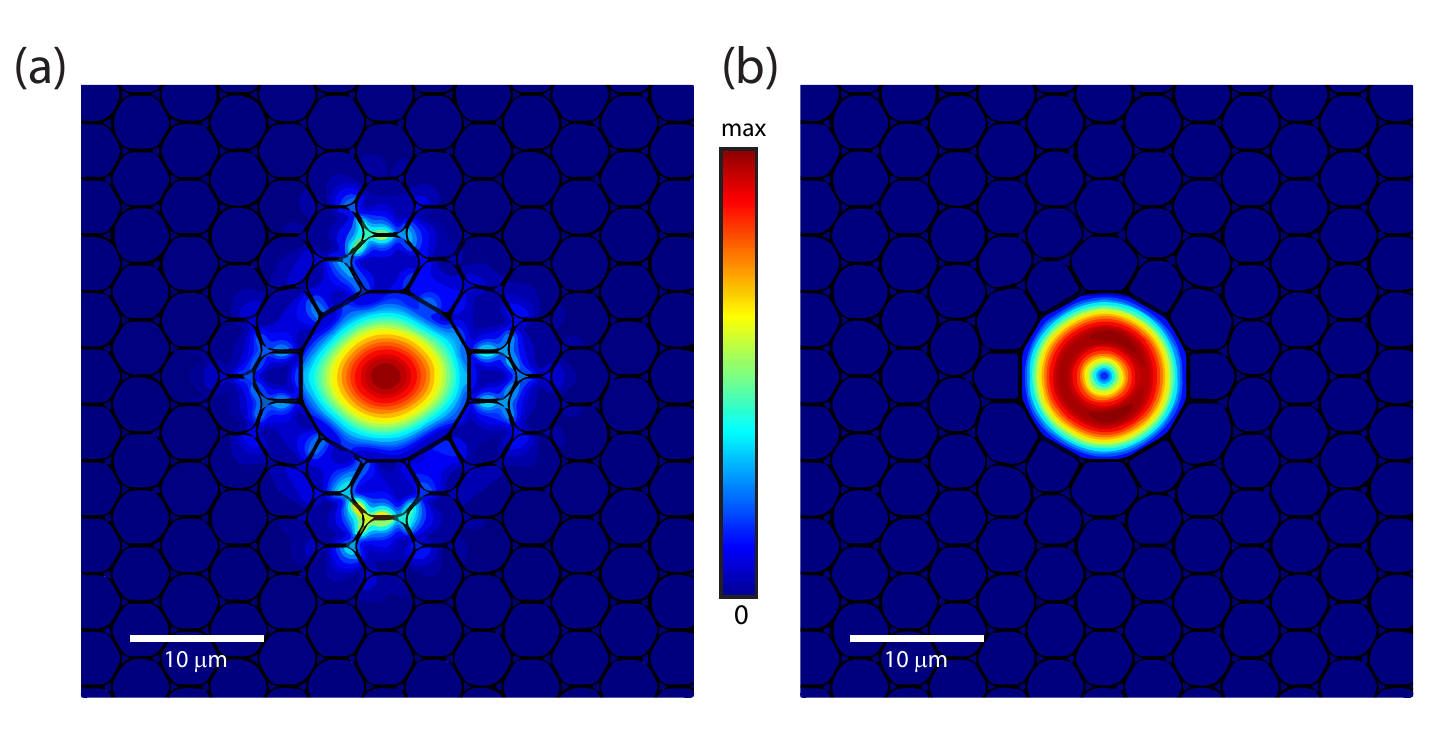}}
\caption{Numerical simulations of the (a) near field intensity profile of the fundamental optical mode and the (b) first excited axial-radial acoustic mode of the air in the core of the hollow-core fiber.}
\label{fig3fullsim}
\end{figure}

To determine whether the discrepancies between the predicted frequencies and coupling strengths derived from our analytical formulation and experiment are due to our simplification of the fiber geometry, numerical simulations were performed. This full-vectorial model contains the complete (HC-1550-02) hollow-core fiber geometry. The air filling all of the voids in the silica fiber matrix is taken to have a mass density of $1.2754$ kg/m$^3$ and a sound speed of 343 m/s.  Using the parameters of Ref. \cite{Aghaie2013} and with additional discussions with these authors, we accurately specify the hollow-core fiber geometry. Using COMSOL multiphysics solver, we solve for the fundamental optical mode (Fig. \ref{fig3fullsim}(a)) as well as for the first excited axial radial (AR1) acoustic mode of the 2D hollow-core geometry (Fig. \ref{fig3fullsim}(b)). The simulated frequency is 36 MHz and the forward-SBS gain is calculated and given by $G^{AR1}=7\times10^{-4}$ W$^{-1}$m$^{-1}$.  This simulated gain agrees well with the measured value, which indicates that the silica/air cladding and the exact non-circular shape of the core contributes to the coupling strength.  Nonetheless, the simple analytical model is useful for simple estimates of the coupling strengths and frequencies for arbitrary gases.

While these studies have focused on the properties of forward-SBS in air, it is also interesting to consider the prospects for enhancement of this effect by filling the fiber core with different atomic vapors. To appreciate the potential for enhancement, it is important to note that forward-SBS gain is proportional to the index of refraction of the Brillouin-active gas. For instance, the refractive index increases sharply for wavelengths near one or more absorption resonance of an atomic vapor.  While high dispersion is typically accompanied by high absorption, this is not always the case. For instance, strong interactions with atomic resonances can be achieved with low optical absorption in the case of electromagnetically induced transparency (EIT) \cite{Boyd2001}.  Since EIT has been demonstrated in hollow-core fibers filled with rubidium \cite{Ghosh2006,Bajcsy2009} as well as with cesium \cite{Sprague2013} vapors, the study of gas mediated Brillouin coupling could give rise to some intriguing new dynamics.  Combining these methods with the gas-mediated Brillouin interactions described here, it may be possible to dramatically enhance and engineer new forms of forward-SBS in gas-filled hollow-core fibers.

In conclusion, we have observed a new form of optomechanical coupling in air-filled hollow-core fibers. Stimulated forward Brillouin scattering is identified at a single 35-MHz resonance, corresponding to the first excited axial radial acoustic mode in the air-filled core.  A simple analytical model is developed, and refined by numerical simulations, which accurately predicts the coupling strength and frequency of the interaction.  This new type of optomechanical interaction is highly tailorable and could allow for forward Brillouin spectroscopy in any low-index gas and for sensing applications.  Finally, SBS in air-filled HC-PBF will present a power and noise limitation for certain applications.

\section*{Funding Information}
The authors acknowledge support from Yale University.

\maketitle
\newcommand{\beginsupplement}{%
        \setcounter{table}{0}
        \renewcommand{\thetable}{S\arabic{table}}%
        \setcounter{figure}{0}
        \renewcommand{\thefigure}{S\arabic{figure}}%
     }

\beginsupplement
\section{Supplement 1}
 
\subsection{Analytical model}
The forward Brillouin gain and resonance frequency of air in a hollow-core fiber can be predicted given the properties of air and the geometry of the fiber. In particular, the forward-SBS gain can be calculated if the optical and acoustic eigenmodes can be determined for the hollow-core fiber geometry. The Brillouin gain $G^{kl}$ is proportional to a force, $\mathbf{f}^{kl}$, for a pump electric field mode with polarization state $k$ (defined as either $s$ or $p$) and a Stokes mode with polarization state $l$. Unlike in solids, pressure waves do not produce birefringence in typical gases (e.g. nonchiral) and therefore $f^{kl}=0$ when $k\neq l$.  Using $\parallel$ to represent $k=l$ and $\perp$ to represent $k\neq l$, and following Refs. \cite{qiu2013stimulated,RakichOE2,Renninger2016}, the Brillouin gain for gases can therefore be written as:
\begin{equation}
G^{\parallel}= \frac{2\omega \,Q}{\Omega^2v_g^2}\frac{|\langle \mathbf{f}^{\parallel},\mathbf{u} \rangle|^2}{\langle \mathbf{E},\epsilon \mathbf{E}\rangle^2\langle \mathbf{u},\rho \mathbf{u}\rangle}\quad \text{and} \quad G^{\perp}=0.
\label{bgaineq}
\end{equation}
$\langle \mathbf{A},\mathbf{B}\rangle$ is a vector inner product, defined as $\langle \mathbf{A},\mathbf{B}\rangle\equiv\int\mathbf{A}\cdot \mathbf{B}^* \,dxdy$, $\epsilon$ is the transverse dielectric distribution of the fiber, $\rho$ is the transverse mass density distribution, $\omega$ is the angular frequency of the Stokes wave, $\Omega$ is the acoustic frequency, $Q$ is the quality factor, $v_g$ is the optical group velocity, $\mathbf{E}$ is the time varying electric field of the optical mode, and $\mathbf{u}$ is the acoustic displacement vector for the acoustic mode.

We assume a simple rigid hollow cylinder with the diameter of the HC-PBF.  The equation for the displacement field, $\mathbf{u}$, given velocity, $v$, is represented as
\begin{equation}
\frac{\partial^2\mathbf{u}}{\partial t^2}=v^2\nabla\nabla\cdot\mathbf{u}.
\label{eqdisp}
\end{equation}\
Because the optical mode is radially symmetric, only the radially symmetric axial-radial acoustic modes will have nonzero coupling.  Assuming that the displacement field oscillates at frequency $\Omega$, the relevant equation for the displacement field is then
\begin{equation}
r^2\frac{\partial^2 u_r}{\partial r^2}+r\frac{\partial u_r}{\partial r}+ \left(r^2\frac{\Omega^2}{v^2}-1\right)u_r=0,
\label{radeqdisp}
\end{equation}
where we have neglected the small contribution from $u_z$.  This is the equation for the Bessel function of the first kind $J_1$.  Therefore, the relevant acoustic modes, $\mathbf{u}_i$, are given by $\mathbf{u}_i=u_o J_1(\frac{\Omega_i}{v}r)\hat{r}$, where $u_o$ is a normalization constant.  The frequencies ($\Omega_i$) are determined by requiring the displacement to be zero at the core of the fiber ($\mathbf{u}=0$ when $r=R$).  This boundary condition gives $\Omega_i=X_{1i}\frac{v}{R}$, where $X_{1i}$ is the $i$th zero of $J_1$.  The coupling integrals are then calculated with this acoustic profile and assuming a Gaussian electric field profile with a mode-field diameter equal to the core radius.  Assuming a dilute gas, the Brillouin gain can then be written as:
\begin{equation}
G^{\parallel}_i=\frac{\pi Q \xi_i n_g^2(n^2-1)^2}{\lambda X_{1i}^2 v^2 n^4 c \rho R^2}\quad \text{and} \quad G^{\perp}_i=0,
\label{ideagasG2}
\end{equation}
where $n$ is the refractive index, $n_g$ is the group index, and $\xi_i$ is a nondimensional coupling constant defined as $\frac{R^4\left(\int\frac{\partial \mathbf{E}\cdot\mathbf{E}}{\partial r}\cdot \mathbf{u_i}dA_R\right)^2}{\int \mathbf{u_i}\cdot\mathbf{u_i}dA_R \left(\int\mathbf{E}\cdot\mathbf{E}dA\right)^2}$, where $dA_R$ represents an area integration over the core and the first five values are given by $\xi=\{1.085,.017,.014,.00832,.00548\}$.  The gain for the first five resonances is plotted in Fig. 2 of the main text along with the optical field and force distributions. Calculations assume an inner diameter of $5.65$ $\mu$m, $Q=8.75$ (from the experimentally measured linewidth), $\lambda=1550$ nm, $n=1.0003$ for air, an air density of $1.2754$ kg/m$^3$, and an air pressure-wave speed of 343 m/s.

\subsection{Experimental procedure and analysis}

\begin{figure}[htb]
\centerline{
\includegraphics[width=8.0cm]{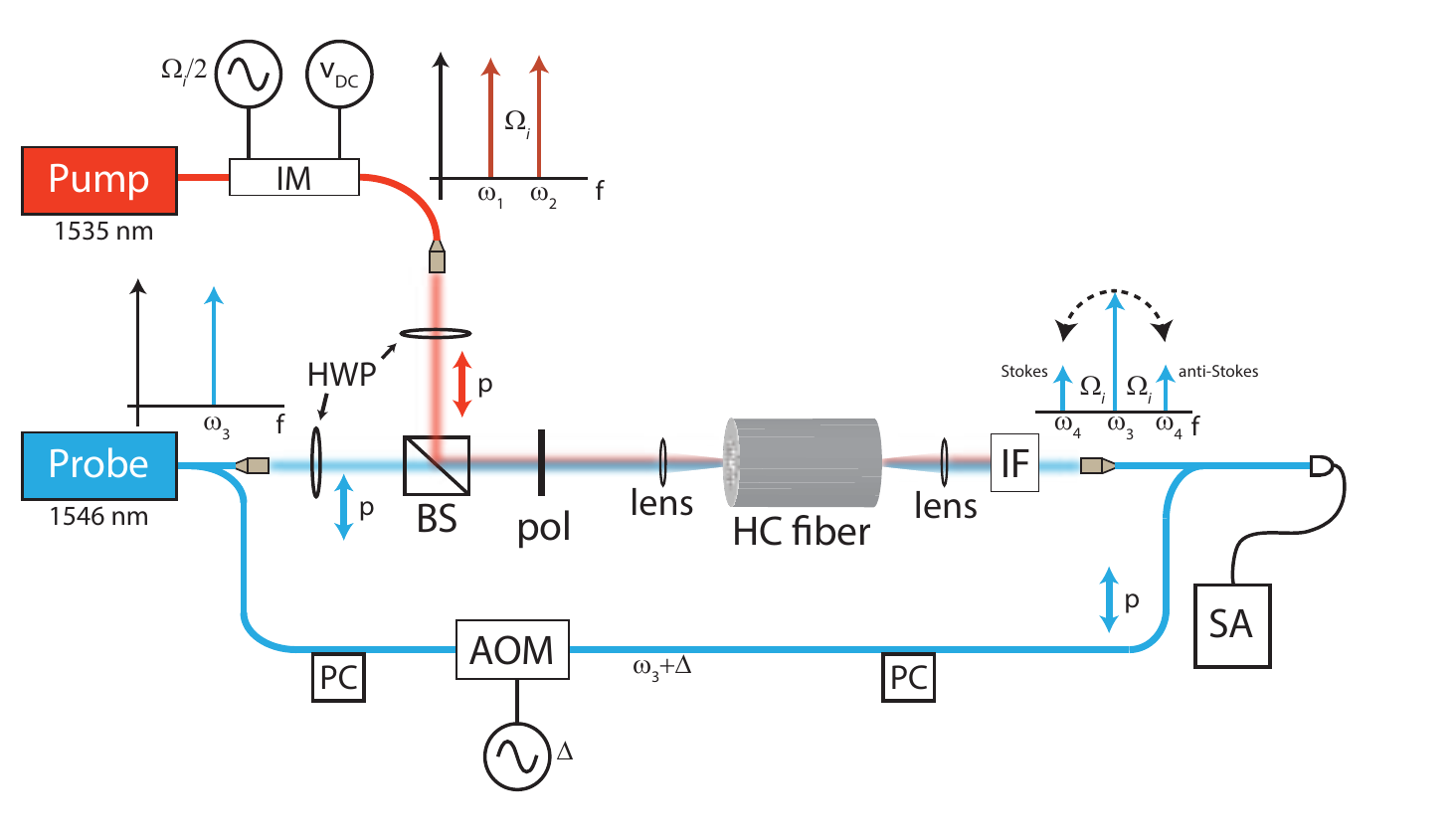}}
\caption{Experimental apparatus: IM, intensity modulator; HWP, half-wave plate; BS, 50:50 beamsplitter; HC fiber, hollow-core fiber; AOM, acousto-optic modulator; pol, polarizer; PC, polarization controller; IF, interference filter; and SA, spectrum analyzer. $\Omega_i$ is the acoustic resonance frequency.}
\label{fig1sch}
\end{figure}

Experimentally, we exploit the broadband nature of forward Brillouin scattering in order to sensitively measure the frequencies and coupling strengths of the interaction \cite{Tang1,Fellegara1997,shin2013,Renninger2016}. Phase matching for the forward-SBS process requires the phase velocity of sound to equal the group velocity of light.  Because the group velocity can have little variation over a large frequency window in fiber, two different wavelength lasers can interact coherently with the same acoustic resonances.  This aspect facilitates the two-color experimental apparatus used in this work (Fig. \ref{fig1sch}).  The two pump-fields (red), $\omega_1$ and $\omega_2$, are synthesized from a monochromatic laser  ($\lambda=1535$ nm) using an intensity modulator.  The modulation frequency is swept through the Brillouin-active resonances producing a resonant excitation of phonons in the HC-PBF. A continuous-wave probe beam (blue), $\omega_3$, is simultaneously injected into the fiber ($\lambda=1546$ nm) with a 50:50 beam splitter, to permit detection of the excited phonons.  Through this process, a new frequency ($\omega_4$) is detected at the output of the fiber.  The pump (red) tones are filtered out before detection.  Finally, a reference arm derived from the probe is frequency shifted with an acousto-optic modulator (AOM) in order to uniquely measure the Stokes and anti-Stokes sidebands \cite{shin2013}.  The input frequencies ($\omega_1-\omega_3$) are aligned to a common polarization with half-wave plates and a common polarizer.
The new output frequency, $\omega_4$, is then polarization analyzed through control of the reference arm polarization to separately measure intra-polarization scattering at the common input polarization and inter-polarization scattering at the orthogonal direction.

In accordance with analytical predictions, intra-polarization scattering measurements reveal a single 4-MHz linewidth peak at 35 MHz (Fig. 4 of the main text).  A peak is not observed through measurements of inter-polarization scattering, which is consistent with Brillouin scattering in a gas.  In addition, multiple narrower peaks corresponding to elastic modes in the silica matrix (see Ref. \cite{Renninger2016}) are observed at various frequencies in both polarizations.  The electronic Kerr nonlinearity contributes a nonzero background at all frequencies, through a four-wave mixing process.  The nonlinear Brillouin gain, $G$, can be determined with the known powers, fiber length, and coupling factor.  When coupled with the AOM shifted reference signal, the measured heterodyne signal is given by $P_{signal}=2\eta\sqrt{P_{4}P_{AOM}}=L G \eta\sqrt{P_{1}P_{2}P_{3}P_{AOM}}$, where $P_{1}$ and $P_{2}$ are the two pump powers (red in Fig. \ref{fig1sch}), $P_{3}$ is the probe power (blue in Fig. \ref{fig1sch}), and $P_{AOM}$ is the optical power in the AOM arm; here $\eta$ is a unitless scale-factor that accounts for the experimental losses produced by numerous fiber components in the optical path between the fiber segment and the detector.  This coefficient was obtained through careful calibration of this apparatus using reference signals of known powers.  Given $P_{1}=P_{2}=48$ mW, $P_{3}=12.8$ mW, $P_{AOM}=0.34$ mW, $L=1.61$ m and $\eta=6.7\%$, we find that the peak power of the 35-MHz peak is given by 15 nW.  The four-wave mixing gain coefficient can be calculated from the frequency independent background signal (\cite{shin2013}) and is  measured here as $\gamma=1.6\times10^{-4}$ (Wm)$^{-1}$.  The forward-SBS gain after subtracting this contribution from Kerr four-wave mixing is then given by $G=9\times10^{-4}$ (Wm)$^{-1}$.  The Brillouin polarization, linewidth, frequency, and gain agree with theoretical predictions.

\subsection{Spontaneous scattering and noise}

The stimulated Brillouin coefficient can be used to estimate spontaneous forward Brillouin scattering, a similar process in which a single pump tone scatters from thermally excited and phase-matched phonons \cite{Kharel2016}.  The total spontaneously generated optical power produced by the Brillouin active phonon mode is given by
\begin{equation}
 P_{s} \cong P_{as} \equiv \int S(\Omega)d\Omega \cong G P_{pump} L\frac{\pi}{2}\frac{c}{\lambda}\frac{ k_B T}{Q}.
\label{stimtospon}
\end{equation}
Here, $P_{s}$($P_{as}$) represents the total spontaneously scattered Stokes (anti-Stokes) power, integrated over the power spectral density ($S(\Omega)$), $G$ is the stimulated gain coefficient, $P_{pump}$ is the pump power, $L$ is waveguide length, $k_B$ is the Boltzman constant, and $T$ is temperature.  For example for forward-SBS in air in the core of the hollow-core fiber, we can use Eq. \ref{stimtospon} to estimate the spontaneous optical scattering rate as follows: considering $G=9\times10^{-4}$ mW$^{-1}$ and $Q_i=8.75$, with $P_{pump}=100$ mW, $L=10$ m, $T=298$ K, and $\lambda=1550$ nm, the total spontaneously generated power scattered by this phonon mode is given by $P_{s}=P_{as}=130$ pW.

\end{document}